 \documentclass[final,3p,times]{elsarticle}

\usepackage{amsmath}
\usepackage{amssymb}
\usepackage{subfigure}
\usepackage[colorlinks = true,
            linkcolor = blue,
            urlcolor  = blue,
            citecolor = red
             ]{hyperref}
\usepackage{tikz}
\usepackage{algorithm}
\usepackage{algpseudocode}
\usepackage[pagewise]{lineno}

\journal{the Journal of the Acoustical Society of America Express Letters}

\begin{document}
\begin{frontmatter}



\title{Phased Microphone Array for Sound Source Localization with Deep Learning}

\author[SJTU-SAA]{Wei MA\corref{cor1}}
\ead{mawei@sjtu.edu.cn}
\cortext[cor1]{Corresponding author}
\author[KeyGo]{Xun LIU\corref{cor2}}
\ead{ae1905kaka@gmail.com}
\cortext[cor2]{Two authors contributed equally to this work}
\address[SJTU-SAA]{School of Aeronautics and Astronautics, Shanghai Jiao Tong University, Shanghai, PR China}
\address[KeyGo]{Shanghai KeyGo Technology Company Limited, Shanghai, PR China}


\begin{abstract}
To phased microphone array for sound source localization, algorithm with both high computational efficiency and high precision is a persistent pursuit.
In this paper convolutional neural network (CNN) a kind of deep learning is preliminarily applied as a new algorithm.
At high frequency CNN can reconstruct the sound localizations with excellent spatial resolution as good as DAMAS, within a very short time as short as conventional beamforming.
This exciting result means that CNN perfectly finds source distribution directly from cross-spectral matrix without given propagation function in advance, and thus CNN deserves to be further explored as a new algorithm.
%
%
%
%
%
%

\end{abstract}


\end{frontmatter}


\section{Introduction}\label{sec:Introduction}
In recent years with the development of society, the awareness of the impact of noise on health has increased significantly, environmental comfort has been becoming more and more important, and consequently acoustic source localization has been increasingly critical in noise diagnosis.
Nowadays phased microphone array has become a standard technique for acoustic source localization.
In the post-processing, the main two categories of traditional algorithms are beamforming and deconvolution algorithms.

Beamforming algorithms construct a dirty map of source distributions from array microphone pressure signals \cite{Johnson-1993}.
Conventional beamforming is simple and robust, however its main disadvantages include poor spatial resolution particularly at low frequencies and poor dynamic range due to side-lobe effects \cite{Michel-2006}.
For algorithms with better performances, many researchers have proposed some advance beamforming algorithms, such as orthogonal beamforming \cite{Sarradj-2010}, robust adaptive beamforming \cite{HuangXun-2012}, and functional beamforming \cite{Dougherty-2014a}.
Concerning spatial resolution, these advance beamforming algorithms have obvious superiority compared to conventional beamforming, however they are not as good as deconvolution algorithms.

Deconvolution algorithms reconstruct a clean map of source distributions from a dirty map via iterative deconvolution, and thus can significantly improve the spatial resolution. The most famous deconvolution algorithms are DAMAS \cite{Brooks-2004, Brooks-2006}, NNLS \cite{Lawson-1995} and CLEAN-SC \cite{Sijtsma-2007}.
However deconvolution algorithms require a relatively high computational effort compared to conventional beamforming due to the inevitable iterations used in the deconvolution algorithms.
Spectral procedure \cite{Dougherty-2005} and compression computational grid \cite{MaWei-2017a, MaWei-2017b, MaWei-2018a} are used to improve the efficiency of deconvolution algorithms.

There are still two big challenges for phased microphone array.
One is that algorithm with both high computational efficiency and high precision is a persistent pursuit, to improve the ability of real-time display and online analysis.
The other one is that when phased microphone array used in complex flow environment with unknown propagation function, phased microphone array with traditional algorithms loses its accuracy, due to uncertainty in the propagation function used in traditional algorithms when.

At this time deep learning - deep neural networks - is the most attractive data mining tool without any doubt.
Deep learning is a specific kind of machine learning \cite{Goodfellow-2017}.
Machine learning is able to learn from data and find the relationship between input and output data.
Deep learning discovers intricate structure in large data sets by using the back propagation algorithm to indicate how a machine should change its internal parameters that are used to compute the representation in each layer form the representation in the previous layer \cite{LeCun-2015}.
Deep learning has recently achieved spectacular success in many domains such as speech recognition \cite{Dahl-2012}, visual object recognition \cite{Krizhevsky-2012}, astronomy \cite{Hezaveh-2017}, as well as the game of Go \cite{Silver-2017}.

In traditional disciplines, deep learning have also attracted widespread attention and expected to be able to further solve the traditional problems.
For example, deep learning has been used to turbulence modelling in fluid mechanics \cite{LingJ-2016, Kutz-2017}.
%
In these traditional disciplines, deep learning is still strongly challenging the deep-rooted consensus that innovations are inspired from expert-in-the-loop intuition and physically interpretable models, by providing competing predictions and without clear physical interpretation.

%


Inspired by the success of deep learning, in this paper convolutional neural network (CNN) \cite{Goodfellow-2017} a kind of deep learning is applied to phased microphone array for sound source localization as a new algorithm.
CNN uses the mathematical operation convolution in at least one of their layers.
Convolution leverages three important ideas that can help improve a machine learning: sparse interactions, parameter sharing and equivariant representations \cite{Goodfellow-2017}.

This attempt mainly looks forward to making full use of three features of deep learning to overcome the big challenges of phased microphone array introduced above.
The first one is the excellent data learning capabilities.
The second one is its computational speed once trained.
The last one is its potential applications with unknown propagation function.
%

%
The rest of this paper is organized as follows.
Algorithms are presented in Section \ref{sec:algorithms}.
An application is examined in Section \ref{sec:application}.
Finally, discussion and conclusion are presented in Section \ref{sec:discussion}.

\section{Algorithms}\label{sec:algorithms}


\begin{figure}[htbp]
  \centering
  \includegraphics[width=0.65\textwidth]{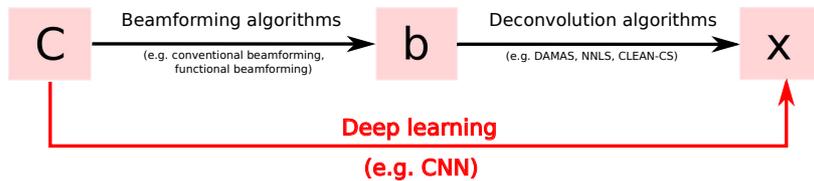}
  \caption{Methods used for source localization}
  \label{fig:methods}
\end{figure}

Fig. 1 of \citet{MaWei-2017b} illustrates a setup with a planar microphone array that contains $M$ microphones and has a diameter of $D$, as well as a two-dimensional region of interest.
Stationary noise sources are located in an $x$-$y$ plane at a distance of $z_0$ from the centre of the microphone array.
The length of the scanning plane is $L$=$2z_0\text{tan}(\alpha/2)$, where $\alpha$ is the opening angle.
The region of interest is divided into $S$=$N\times N$ equidistant points.

In each test case, data from the microphone array are simultaneously acquired.
Cross-spectral matrix (CSM) is then calculated using these simultaneously acquired data from the microphone array.
The acquired data of each microphone are divided into $I$ frames.
Each frame is then converted into frequency bins by Fast Fourier Transform (FFT).
For a given angular frequency $\omega$, CSM is averaged over $I$ blocks
\begin{equation}\label{eq:CSM}
\mathbf{C}(\omega) = \overline{\mathbf{p}(\omega)\mathbf{p}(\omega)^H} = \dfrac{1}{I}\sum_{i=1}^{I}\mathbf{p}_i(\omega)\mathbf{p}_i(\omega)^H
\end{equation}
where $\mathbf{p}(\omega)=[p_1(\omega), p_2(\omega), ..., p_M(\omega)]^T$, $(\cdot)^H$ denotes complex conjugate transpose.
For the sake of brevity, $\omega$ is omitted in the following.
%
The problem of phased microphone arrays for source localization can be expressed as
\begin{equation}\label{eq:fcx}
f(\mathbf{C})= \mathbf{x}
\end{equation}
where $\mathbf{x}$ is the source distribution of power descriptors and
\begin{linenomath*}
\begin{equation}
\mathbf{x}=[q_1^2, ..., q_s^2, ..., q_S^2]^T
\end{equation}
\end{linenomath*}
where $q_s$ is source amplitude in terms of the pressure produced at source point $s$.

Fig. \ref{fig:methods} shows the algorithms deal with Eq. \ref{eq:fcx}, including beamforming algorithms, deconvolution algorithms, and deep learning.

\subsection{Beamforming algorithms}
The conventional beamforming
\begin{linenomath*}
\begin{equation}\label{eq:dirty_map_2}
b(\mathbf{r})=\dfrac{\mathbf{e}(\mathbf{r})^H \mathbf{C} \mathbf{e}(\mathbf{r})}{||\mathbf{e}(\mathbf{r})||^4}
\end{equation}
\end{linenomath*}
where the vector $\mathbf{e}(\mathbf{r})$$\in \mathbb{C}^{M \times 1}$ is the steering vector at $\mathbf{r}$ and
\begin{linenomath*}
\begin{equation}
\mathbf{e}(\mathbf{r})=[e_1(\mathbf{r}),...,e_m(\mathbf{r}),...,e_m(\mathbf{r})]^T
\end{equation}
\end{linenomath*}
The notation of steering vector under monopole point source assumption and in a medium with a uniform flow is \cite{Brooks-2006}
\begin{linenomath*}
\begin{equation}
e_m(\mathbf{r})=\dfrac{||\mathbf{r}-\mathbf{r}_m||}{||\mathbf{r}||}{\text{exp}}\{-j2\pi f/c_0||\mathbf{r}-\mathbf{r}_m||\}
\end{equation}
\end{linenomath*}
where $||\mathbf{r}||$ is the distance from the beamformer focus position to the centre of the array, $||\mathbf{r}-\mathbf{r}_m||$ is the distance from the beamformer focus position to the $m$th microphone (see in Fig. \ref{fig:array_illustration}), $f$ is frequency, and $c_0$ is speed of sound.


\subsection{Deconvolution algorithms}
The sound pressure contribution at microphones can be written as
\begin{linenomath*}
\begin{equation}
\mathbf{p}=\sum_{s=1}^{S}\mathbf{e}(\mathbf{r_s})q_s
\end{equation}
\end{linenomath*}
For incoherent acoustic sources, CSM thus becomes
\begin{equation}\label{eq:CSM_training}
\mathbf{C}=\sum_{s=1}^{S}|q_s|^2\mathbf{e}(\mathbf{r_s})\mathbf{e}(\mathbf{r_s})^H
\end{equation}
The conventional DAS beamforming output can then be written as
\begin{linenomath*}
\begin{equation}\label{eq:dm1}
b(\mathbf{r})=\sum_{s=1}^{S}|q_s|^2 \cdot \dfrac{\mathbf{e}(\mathbf{r})^H
[\mathbf{e}(\mathbf{r}_s)\mathbf{e}(\mathbf{r}_s)^H]\mathbf{e}(\mathbf{r})}{||\mathbf{e}(\mathbf{r})||^4}
=\sum_{s=1}^{S}|q_s|^2 \cdot \dfrac{|\mathbf{e}(\mathbf{r})^H\mathbf{e}(\mathbf{r}_s)|^2}{||\mathbf{e}(\mathbf{r})||^4}
\end{equation}
\end{linenomath*}
For a single unit-power point source, Eq. (\ref{eq:dm1}) is known as point-spread function (PSF) of the array and is defined as
\begin{linenomath*}
\begin{equation}\label{eq:PSF1}
\mathrm{PSF}(\mathbf{r}|\mathbf{r}_s)= \dfrac{\mathbf{e}(\mathbf{r})^H
[\mathbf{e}(\mathbf{r}_s)\mathbf{e}(\mathbf{r}_s)^H]\mathbf{e}(\mathbf{r})}{||\mathbf{e}(\mathbf{r})||^4}
= \dfrac{|\mathbf{e}(\mathbf{r})^H\mathbf{e}(\mathbf{r}_s)|^2}{||\mathbf{e}(\mathbf{r})||^4}
\end{equation}
\end{linenomath*}
and then Eq. (\ref{eq:dm1}) can then be written as
\begin{linenomath*}
\begin{equation}\label{eq:dm2}
b(\mathbf{r})=\sum_{s=1}^{S}|q_s|^2 \cdot \mathrm{PSF}(\mathbf{r}|\mathbf{r}_s)
\end{equation}
\end{linenomath*}
By computing $\mathrm{PSF}(\mathbf{r}|\mathbf{r}_s)$ for all combinations of $(\mathbf{r}|\mathbf{r}_s)$ in discrete grid and arranging each resulting PSF map column-wise in a matrix $\mathbf{A}$, Eq. (\ref{eq:dm2}) could reformulate in matrix notation as
\begin{linenomath*}
\begin{equation}\label{eq:Axb1}
\mathbf{A x=b}
\end{equation}
\end{linenomath*}
where $\mathbf{b}$ contains the beamformer map.
Eq. (\ref{eq:Axb1}) is a system of linear equations.
Notice that  $\mathbf{A} \in \mathbb{R}^{S \times S}$, $\mathbf{x} \in \mathbb{R}^{S \times 1}$, $\mathbf{b} \in \mathbb{R}^{S \times 1}$.

The deconvolution task is to find a source distribution $\mathbf{x}$ for a give dirty map $\mathbf{b}$ and know matrix $\mathbf{A}$.
The constraint is that each component of the vector $\mathbf{x}$ is larger or equal to zero.
In most of the applications the matrix $\mathbf{A}$ is singular, and $\mathbf{b}$ is in the range of $\mathbf{A}$, this means there are very large number of solutions of $\mathbf{x}$ that fulfil Eq. \ref{eq:Axb1}.
The DAMAS algorithm \cite{Brooks-2006} is an iterative algebraic deconvolution method.
In this algorithm, the source distribution is calculated by the solution of Eq. \ref{eq:Axb1} using a Gauss-Seidel-type relaxation.
In each step the constraint is applied that the source strength remains positive.
%

\subsection{Deep learning}
Deep learning is used to reconstruct source distribution from CSM directly.
Thus input tensor is $\mathbf{C} \in \mathbb{C}^{M \times M}$, while output tensor is $\mathbf{x} \in \mathbb{C}^{S \times 1}$.
Keras framework \cite{Chollet-2015} with a Tensorflow backend is used here.

\subsubsection{Networks architecture}
CNN a kind of deep learning is used here.
Variables of parameters and structures of CNN are displayed in Table \ref{tab:para_CNN}.
This CNN model consists of four two-dimensional convolutional layers (Conv2D), two two-dimensional pooling layers (MaxPooling2D), a flatten layer (Flatten) and a regular densely-connected neural networks layer (Dense).
The convolutional layers perform discrete convolution operations on their input.
In each convolutional layer, zero-padding is valid such that the output has the same length as the original input, meanwhile a bias vector is created and added to the outputs.
The output of each convolutional layer is passed to a rectified linear unit (ReLU) filter.
The pooling layer performs a max operation over sub-regions of the extracted feature maps resulting in down sampling by a factor of two.
The flatten layer just flattens the input and does not affect the batch size.
The regular densely-connected neural networks layer gives $S$-dimensional output space using a matrix multiplication and bias addition.


\begin{table}[htbp]
\centering
\caption{Variables of parameters and structures of the convolutional neural networks.
}
\label{tab:para_CNN}
\begin{tabular}{cccccccc}\hline
Layer& Layer & Kernel & Kernel & Stride & Activation& Padding & Output  \\
No. & Type & Number & Size &  &  &   & Size  \\\hline
1  & Conv2D & 64 & 3$\times$3   & 1$\times$1 & ReLU & Yes   & $M\times M \times 64$ \\
2  & Conv2D & 64 & 3$\times$3   & 1$\times$1 & ReLU & Yes    & $M\times M \times 64$\\
3  & MaxPooling2D & - & 2$\times$2 &  2$\times$2& - & No   & $M/2 \times M/2 \times 64$\\
4  & Conv2D & 128 &  3$\times$3  & 1$\times$1 & ReLU & Yes   & $M/2 \times M/2 \times 128$\\
5  & Conv2D & 128 & 3$\times$3   & 1$\times$1 & ReLU & Yes   & $M/2 \times M/2 \times 128$\\
6  & MaxPooling2D & - & 2$\times$2 &  2$\times$2 & - & No   & $M/4 \times M/4 \times 128$ \\
7  & Flatten & - & - &  - & -  &  -& $(M/4*M/4*128) \times 1$  \\
8  & Dense & 1 & $S \times (M/4*M/4*128)$ & - & -& -   & $S\times 1$\\\hline
\end{tabular}
\\$S$, grid number; $M$, microphone number.
\end{table}

\subsubsection{Training strategy}
To train the CNN, a variant of stochastic gradient descent called ADAM is used.
The learning rate is set as $\alpha$=0.001 and the other hyper-parameters of ADAM optimizer to $\beta_1$=.09, $\beta_2$=0.999 and $\epsilon$=10$^{-8}$ as recommended.
The loss function used to train the weights of the networks is the mean squared error.
Metric function is set as the mean-squares of the errors between assigned and predicted values,
\begin{equation}
\text{Metric Function}=\dfrac{1}{S}\sum_{i=1}^{N}(Y_i-\hat{Y_i})^2
\end{equation}
where $Y_i$ and $\hat{Y_i}$ are assigned value and predicted value at $i^{th}$ grid, respectively.

\subsubsection{Training data}
The data used to train the network is obtained by simulating a CSM for a given sound source distribution according to Eq. \ref{eq:CSM_training}.
%

%


\clearpage
\section{Application}\label{sec:application}


In this section synthetic applications are carried out to check the spatial resolution of CNN.

%
The planar array contains 30 simulated microphones and has a diameter $D$ of 0.35 m, as shown in Fig. 2 of \citet{MaWei-2017b}.
In the geometrical setup, the observation plane is parallel to the array plane, and the region of interest is right in front of the array.
The distance between array plane and observation plane $z_0$ is 2.0 m.
The opening angle $\alpha$=45$^{\circ}$.
The computational grid is 15$\times$15 with 225 grid points.
Gaussian white noise is added with a signal-to-noise ratio of 15 dB at the microphone array.
%

For traditional algorithms, diagonal removal is applied on the CSM used for conventional beamforming, while no diagonal removal is applied on the PSF used for DAMAS.
DAMAS is run with 1000 iterations, which appeared to be more than enough for convergence.

For new algorithm, CNN described in previous section has approximately 1.62$\times$10$^6$ trainable parameters according to microphone number $M$=30 and grid number $S$=225.
The data used to train the network is obtained by simulating a CSM with three uniform sound sources randomly distributed in the grid as the given sound distributions.
In this application for a training, 4$\times$10$^4$ numerical data are generated, in which 80\% are used as training data, 10\% are used as validation data, the remaining 10\% are used as test data.
Test data are made sure do not appear in training and validation data.
Number of samples per gradient update is specified as 32, and number of epochs to train the model is specified as 10 which appeared to be more than enough for convergence.
The network training takes around 4 hours on a MacBook Pro with a processor of 2.9 GHz Inter Core i5 for training.

\begin{figure}[htbp]
\centering
  \subfigure[]{
    \label{fig:beamformer_map} 
    \includegraphics[width=0.27\textwidth]{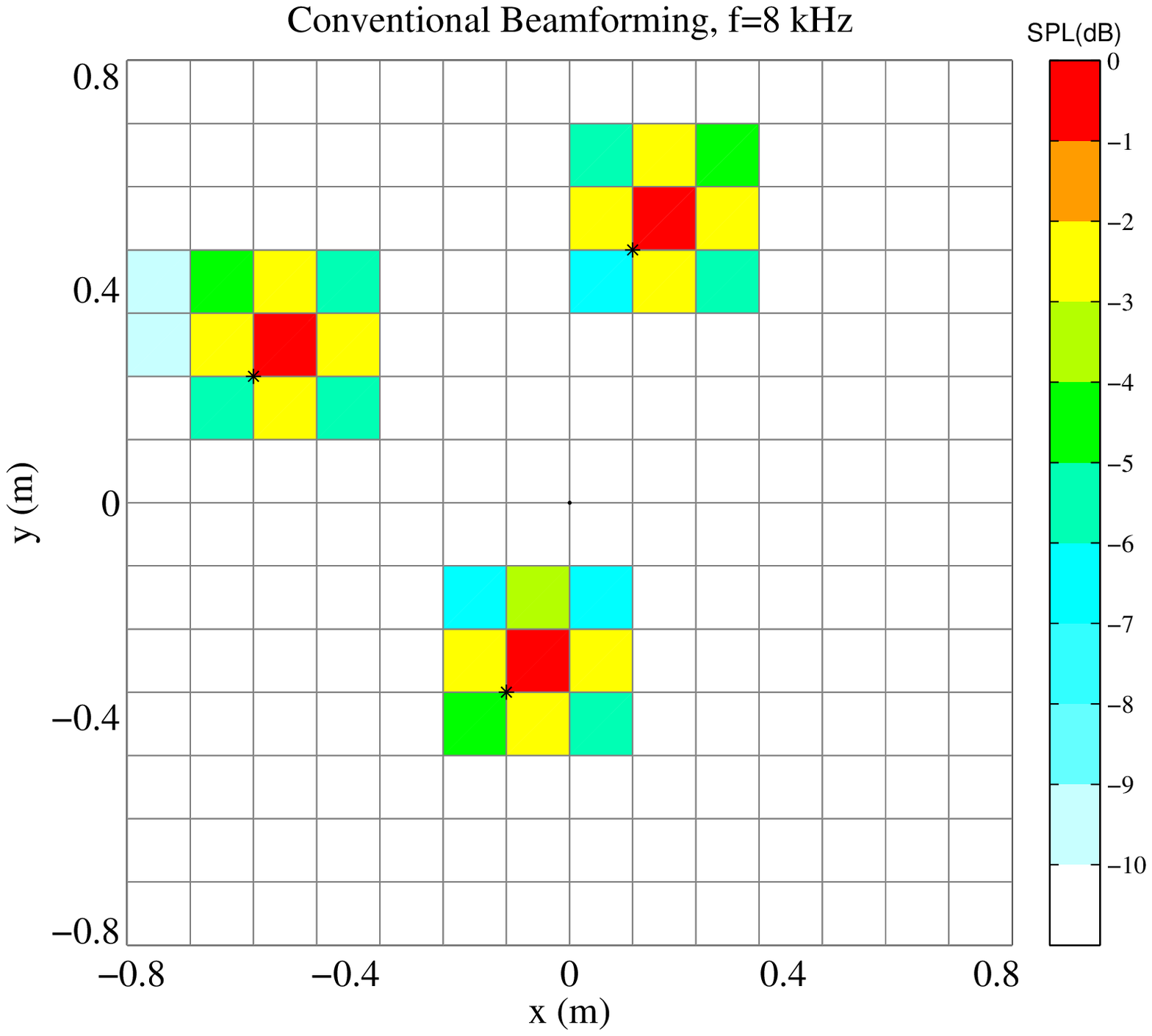}}
  \subfigure[]{
    \label{fig:beamformer_map} 
    \includegraphics[width=0.27\textwidth]{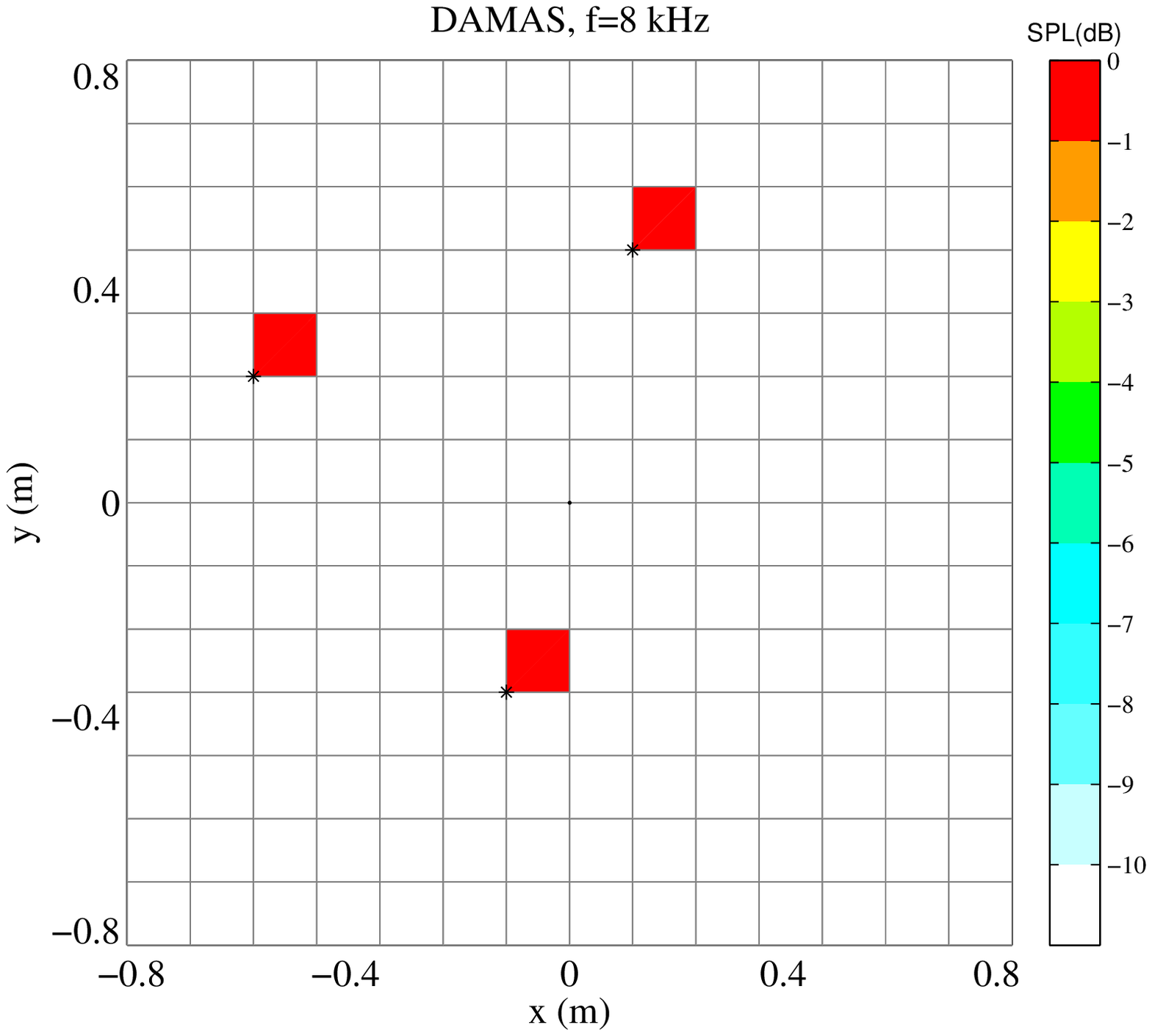}}
   \subfigure[]{
    \label{fig:DAMAS} 
    \includegraphics[width=0.27\textwidth]{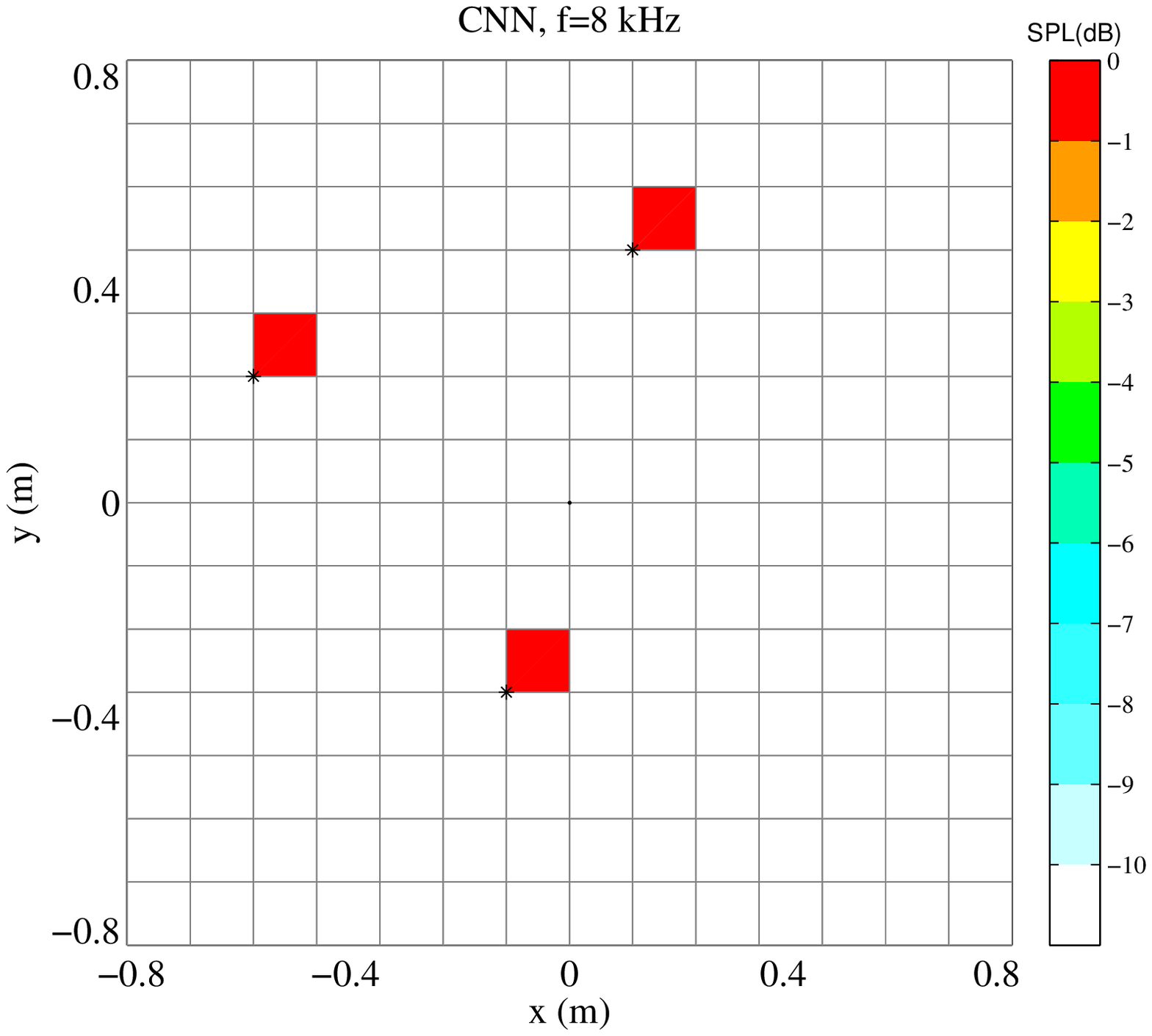}}\\
  \subfigure[]{
    \label{fig:beamformer_map} 
    \includegraphics[width=0.27\textwidth]{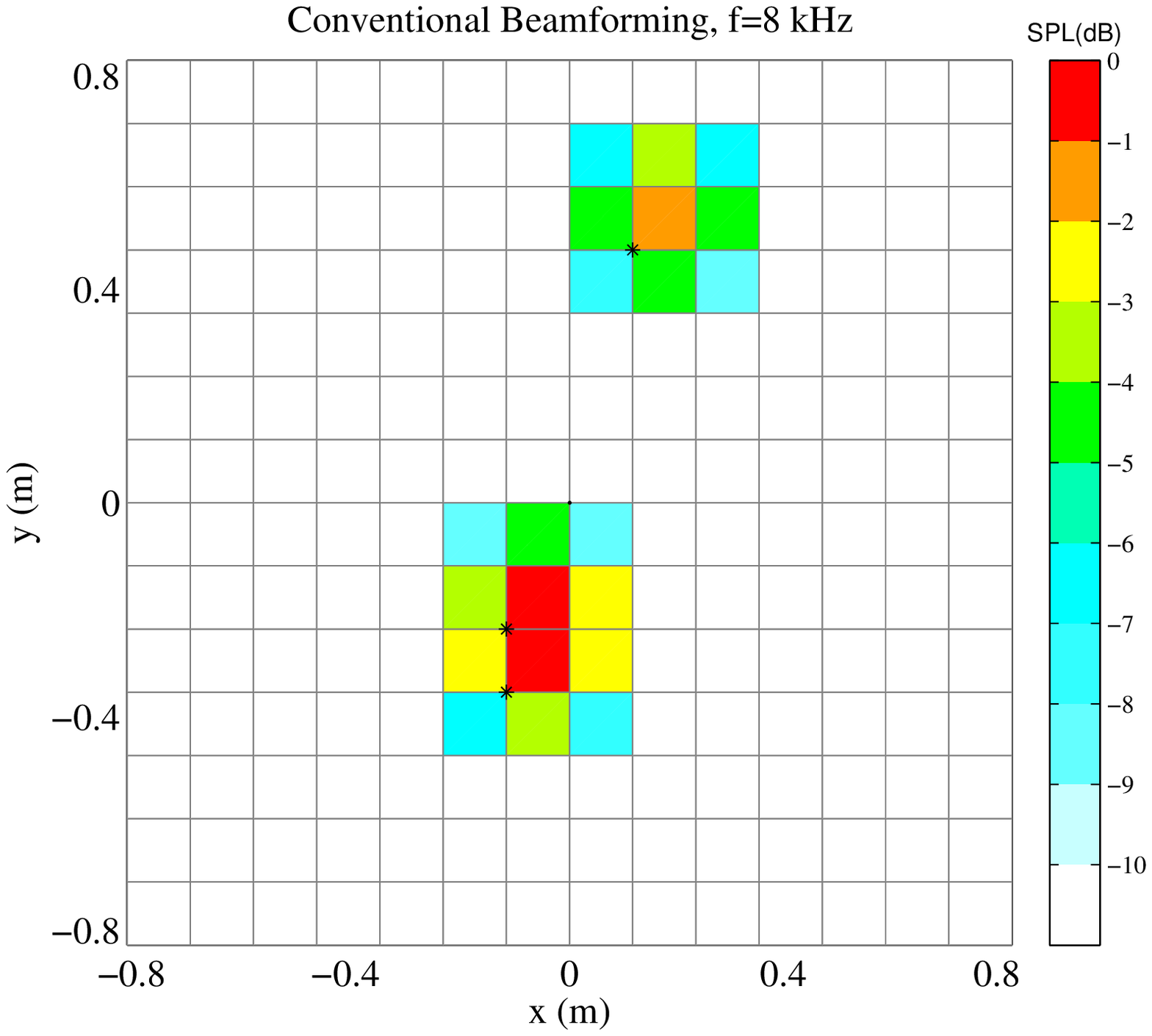}}
  \subfigure[]{
    \label{fig:beamformer_map} 
    \includegraphics[width=0.27\textwidth]{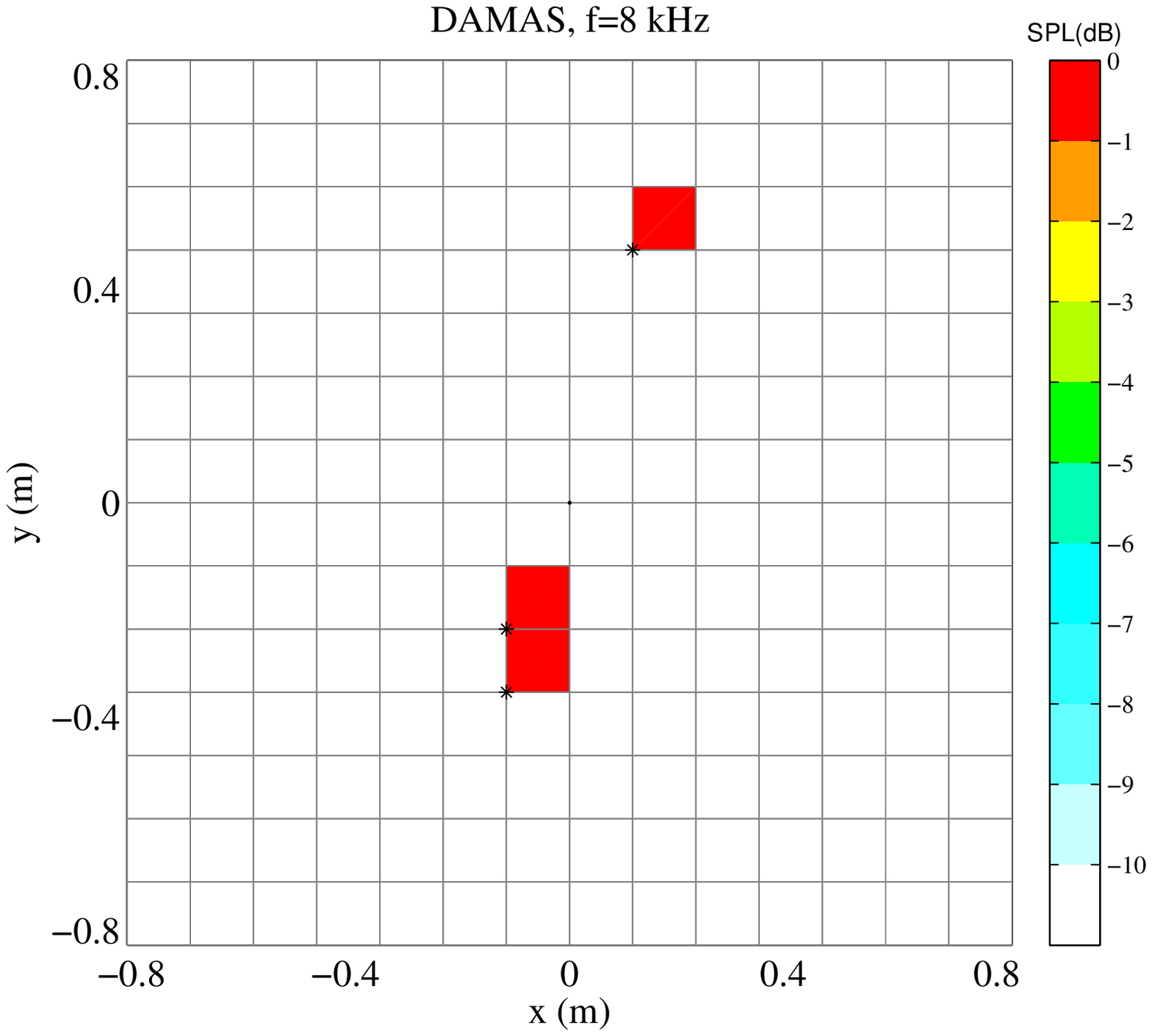}}
   \subfigure[]{
    \label{fig:DAMAS} 
    \includegraphics[width=0.27\textwidth]{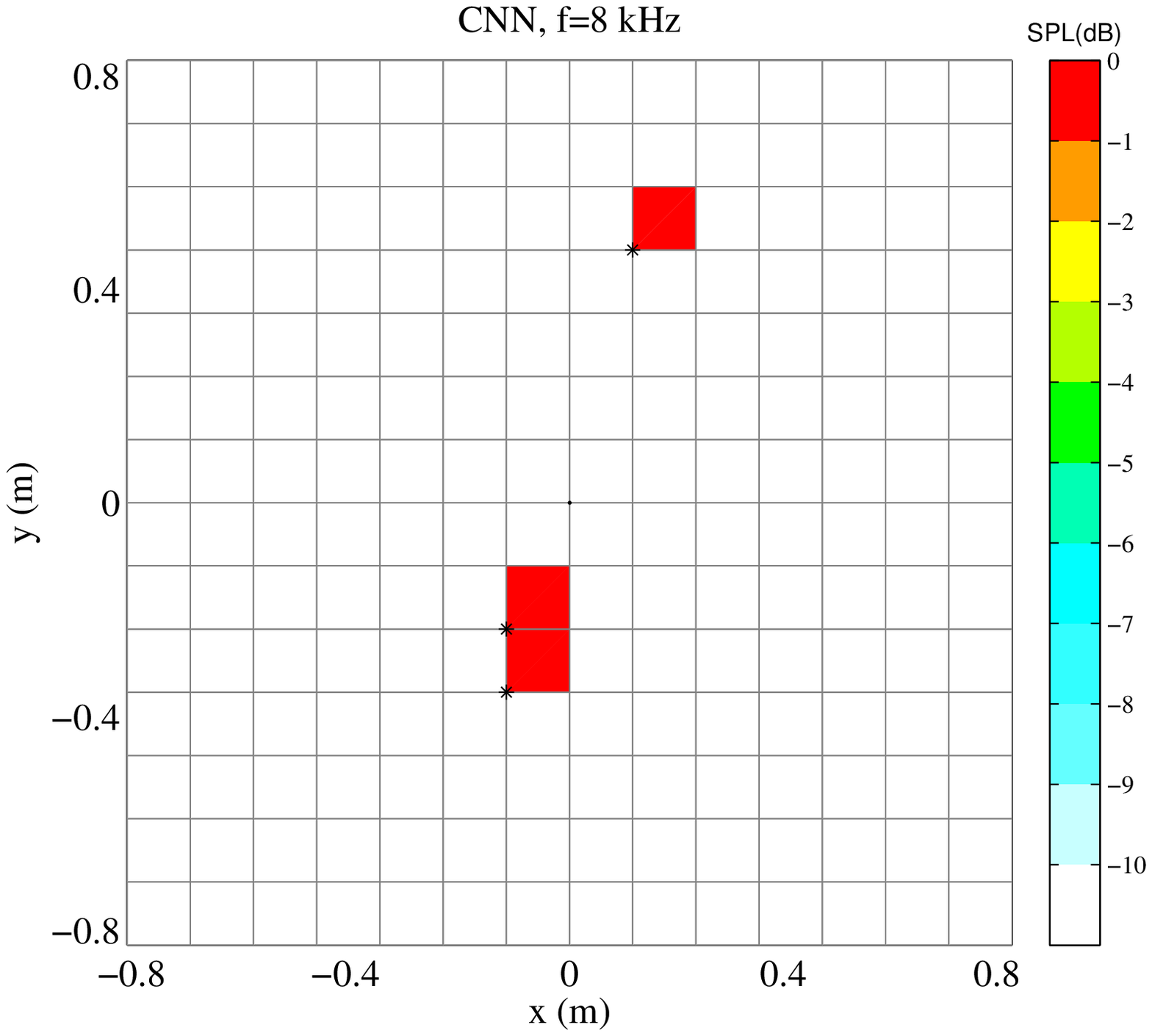}}\\
\caption[]{$f$=8 kHz. Black cross symbols, positions of synthetic point sources. The first line, three dispersed sources far apart; the second line, two of three sources are adjacent. The first column, conventional beamforming; the second column DAMAS; the third column, CNN.}
\label{fig:f8}
\end{figure}

CNN test accuracy with frequency is listed in Table \ref{tab:CNN_accuracy}.
CNN test accuracy at $f$=8 kHz is up to 98\%.
After checking these 2\% incorrect examples, at least one sound source point is located at the edge of grid in most cases.
Results of two given sound distributions at $f$=8 kHz are shown in Fig. \ref{fig:f8}.
In the first given sound distribution, the distances between three sound sources are quite large.
Conventional beamforming can separate these three sources although some side-lobes exist.
Both CNN and DAMAS can reconstruct accurately the source distribution.
In the second given sound distribution, two sources are located on adjacent grids.
Conventional beamforming cannot separate these two adjacent sources.
Both CNN and DAMAS can still reconstruct accurately the source distribution.
This exciting result means that CNN almost perfectly finds source distribution $\mathbf{x}$ from $\mathbf{C}$ without given propagation function in advance.

\begin{figure}[htbp]
\centering
  \subfigure[]{
    \label{fig:beamformer_map} 
    \includegraphics[width=0.27\textwidth]{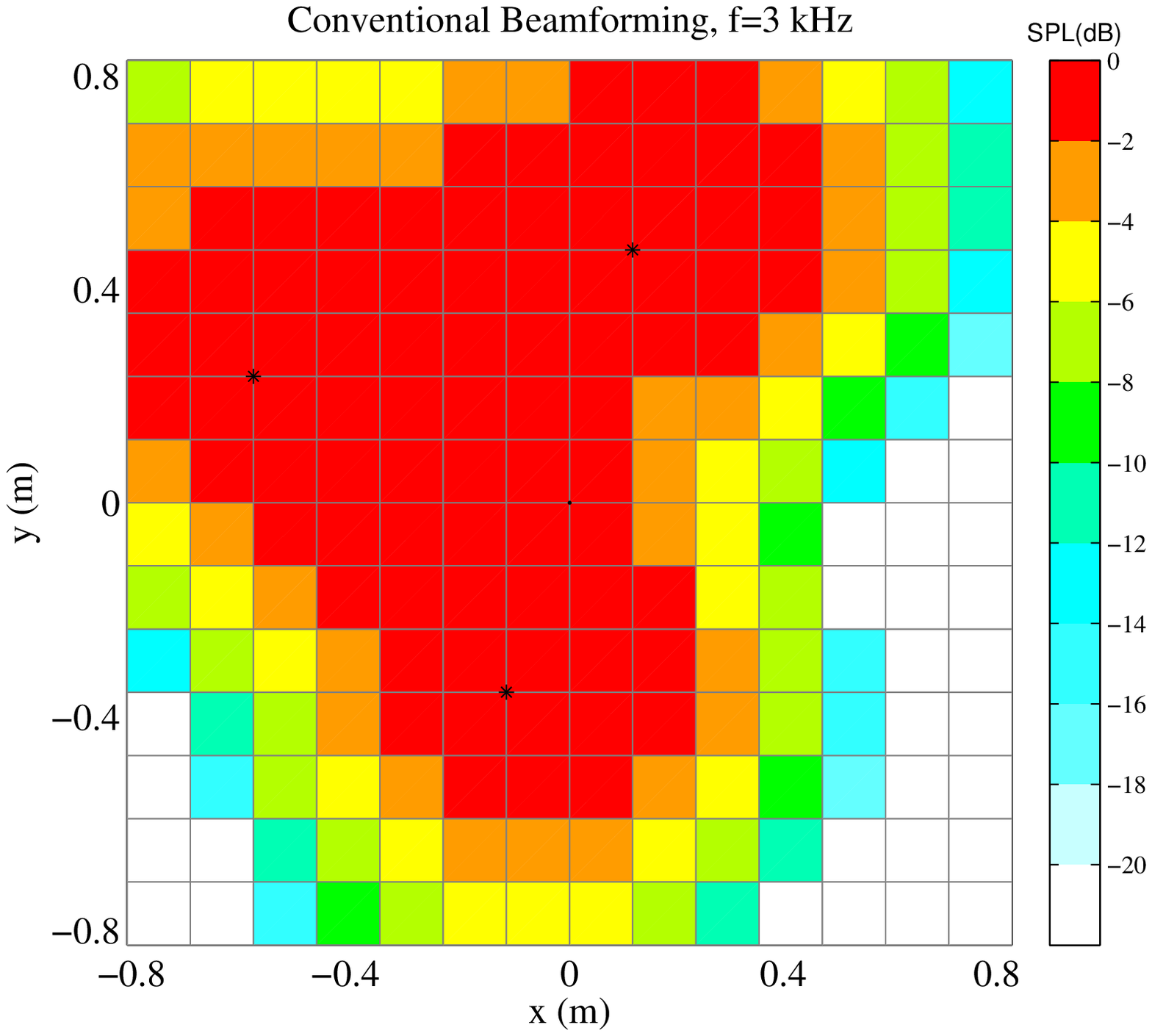}}
  \subfigure[]{
    \label{fig:beamformer_map} 
    \includegraphics[width=0.27\textwidth]{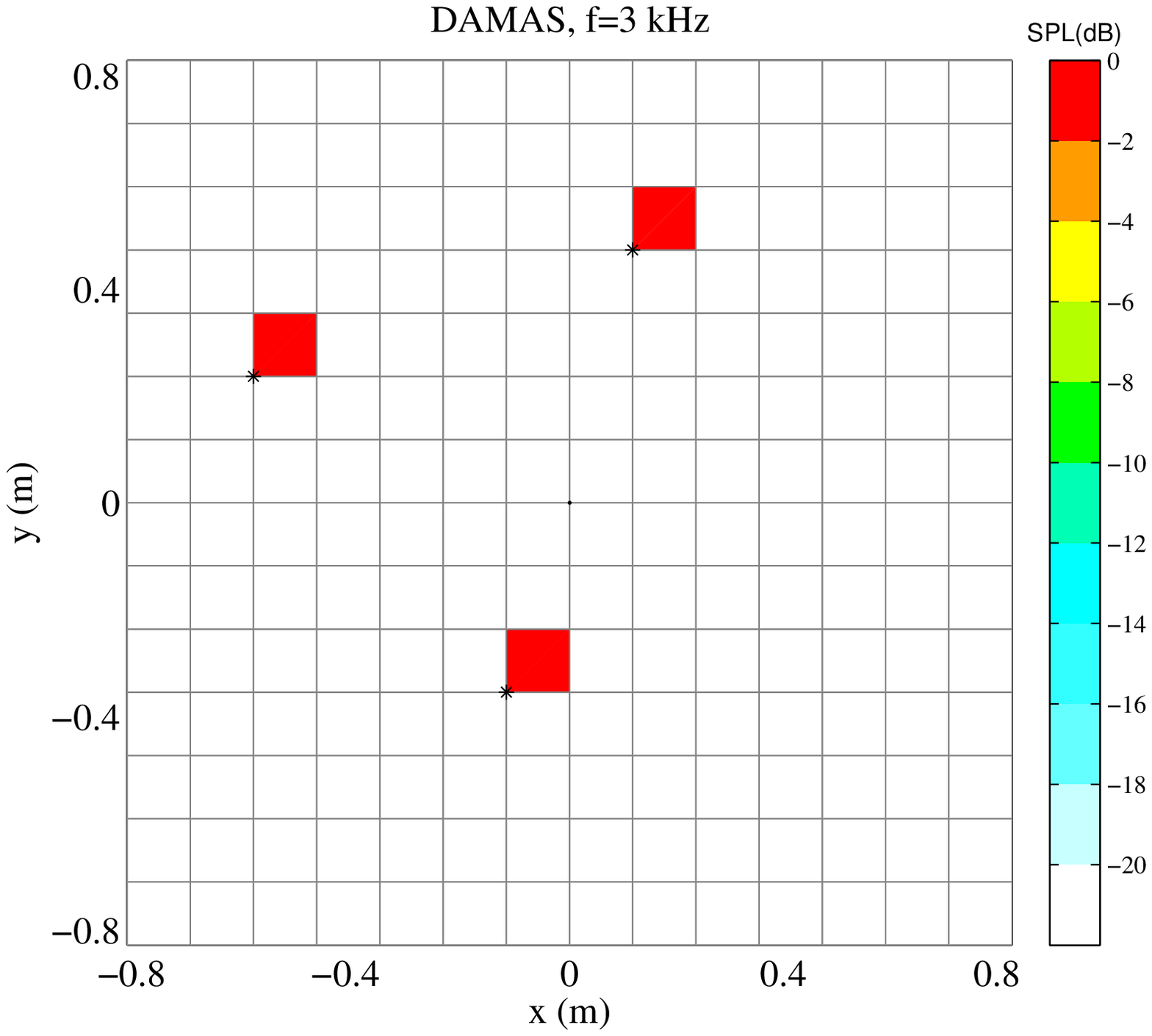}}
   \subfigure[]{
    \label{fig:DAMAS} 
    \includegraphics[width=0.27\textwidth]{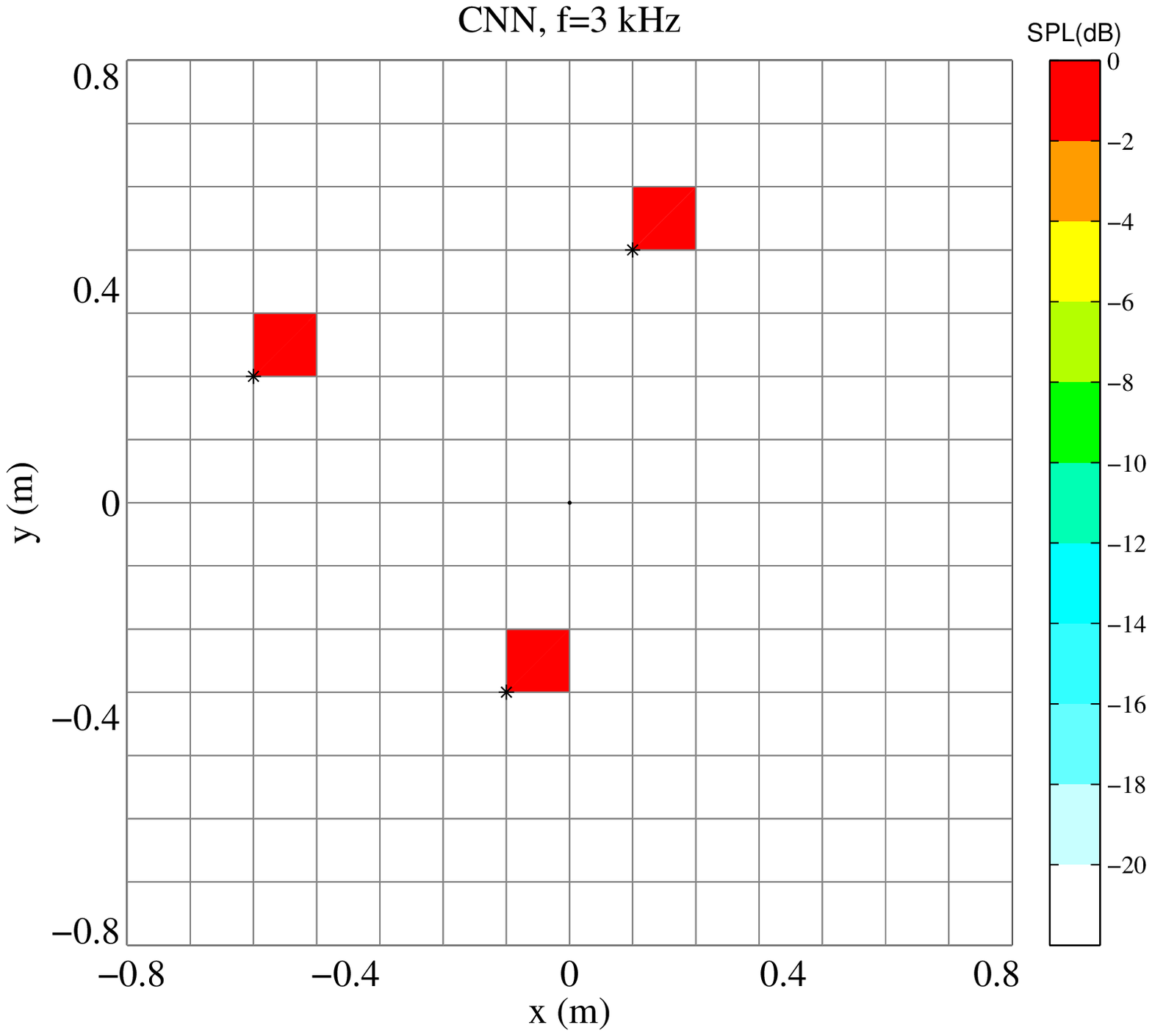}}\\
  \subfigure[]{
    \label{fig:beamformer_map} 
    \includegraphics[width=0.27\textwidth]{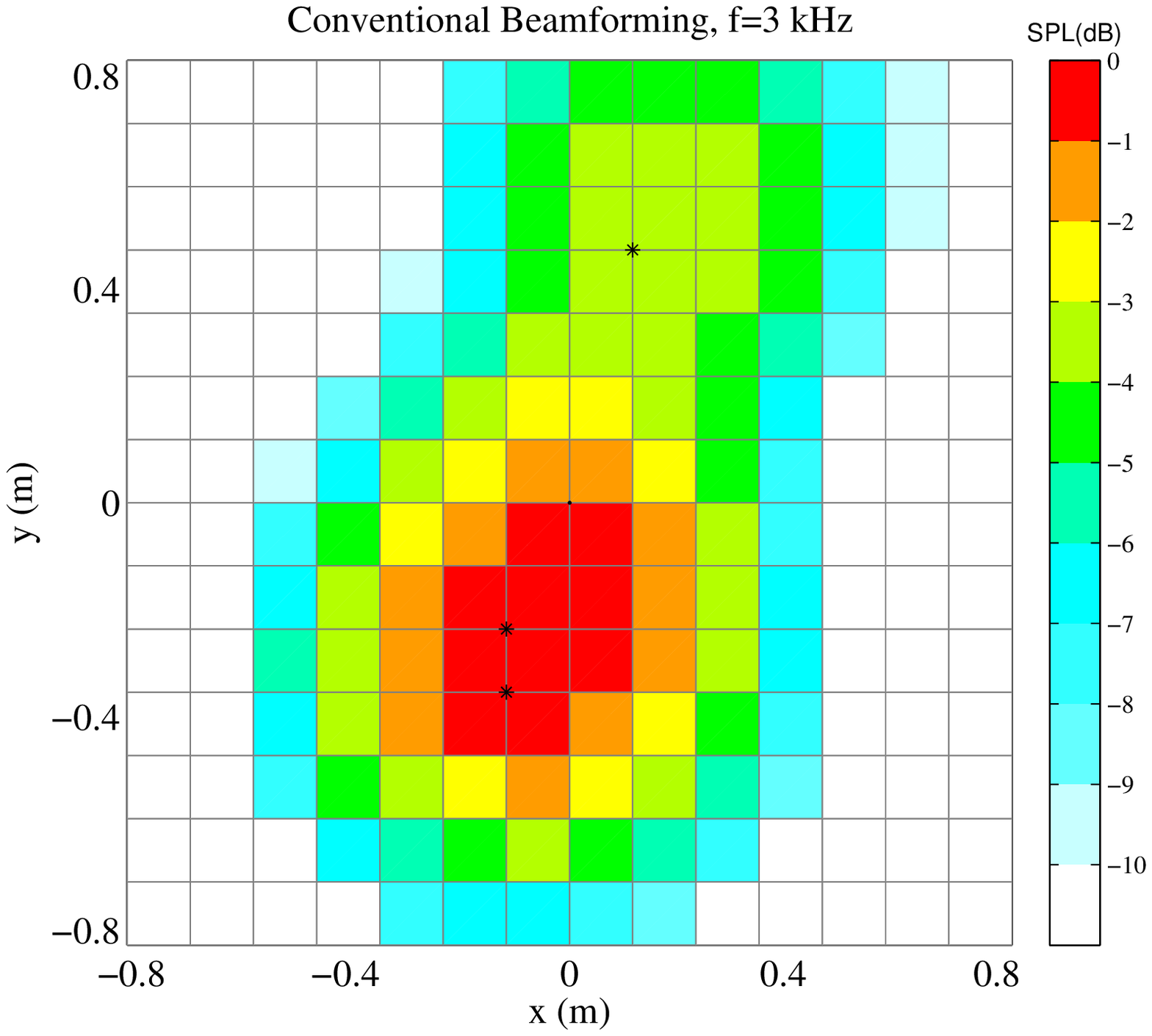}}
  \subfigure[]{
    \label{fig:beamformer_map} 
    \includegraphics[width=0.27\textwidth]{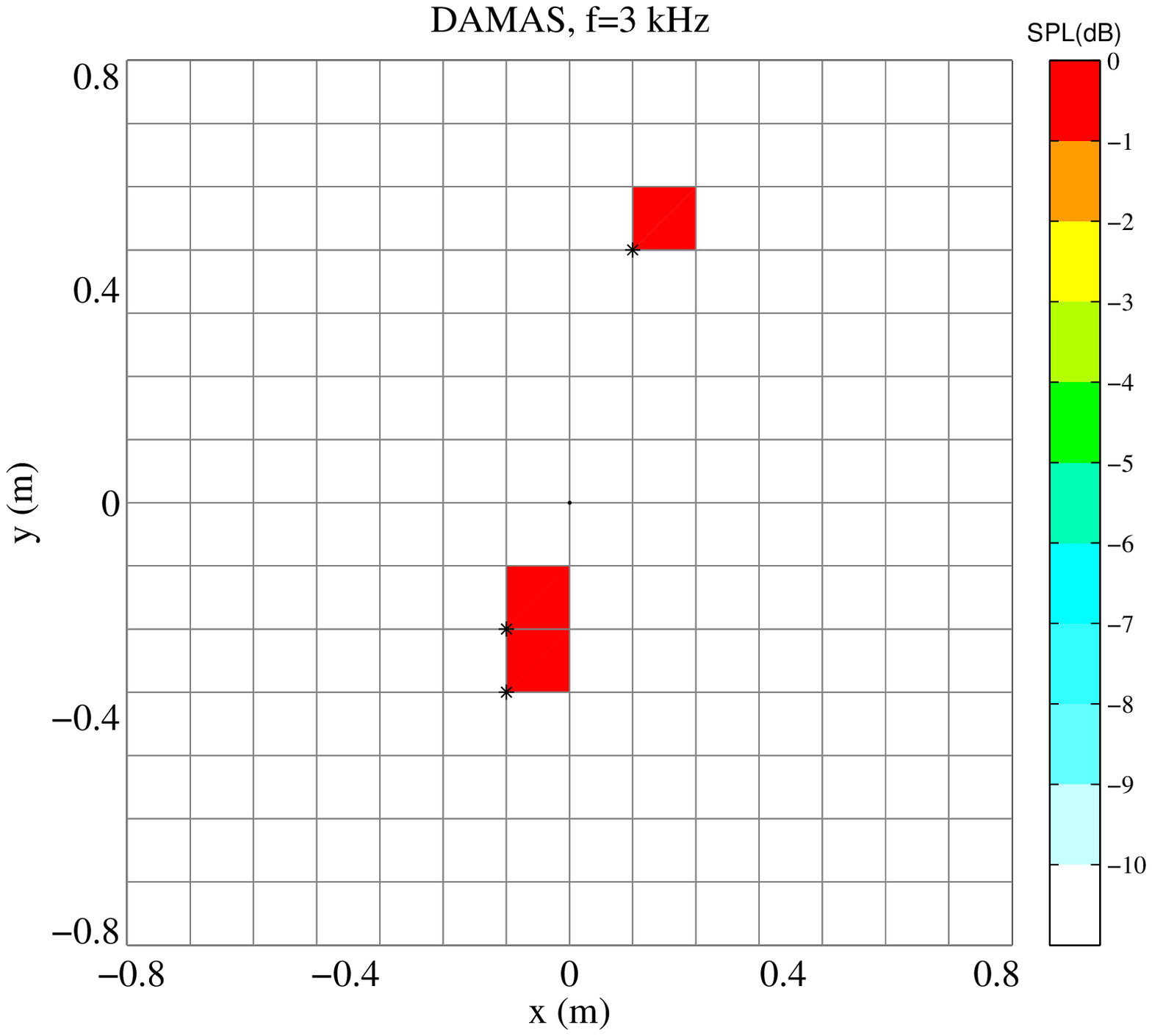}}
   \subfigure[]{
    \label{fig:DAMAS} 
    \includegraphics[width=0.27\textwidth]{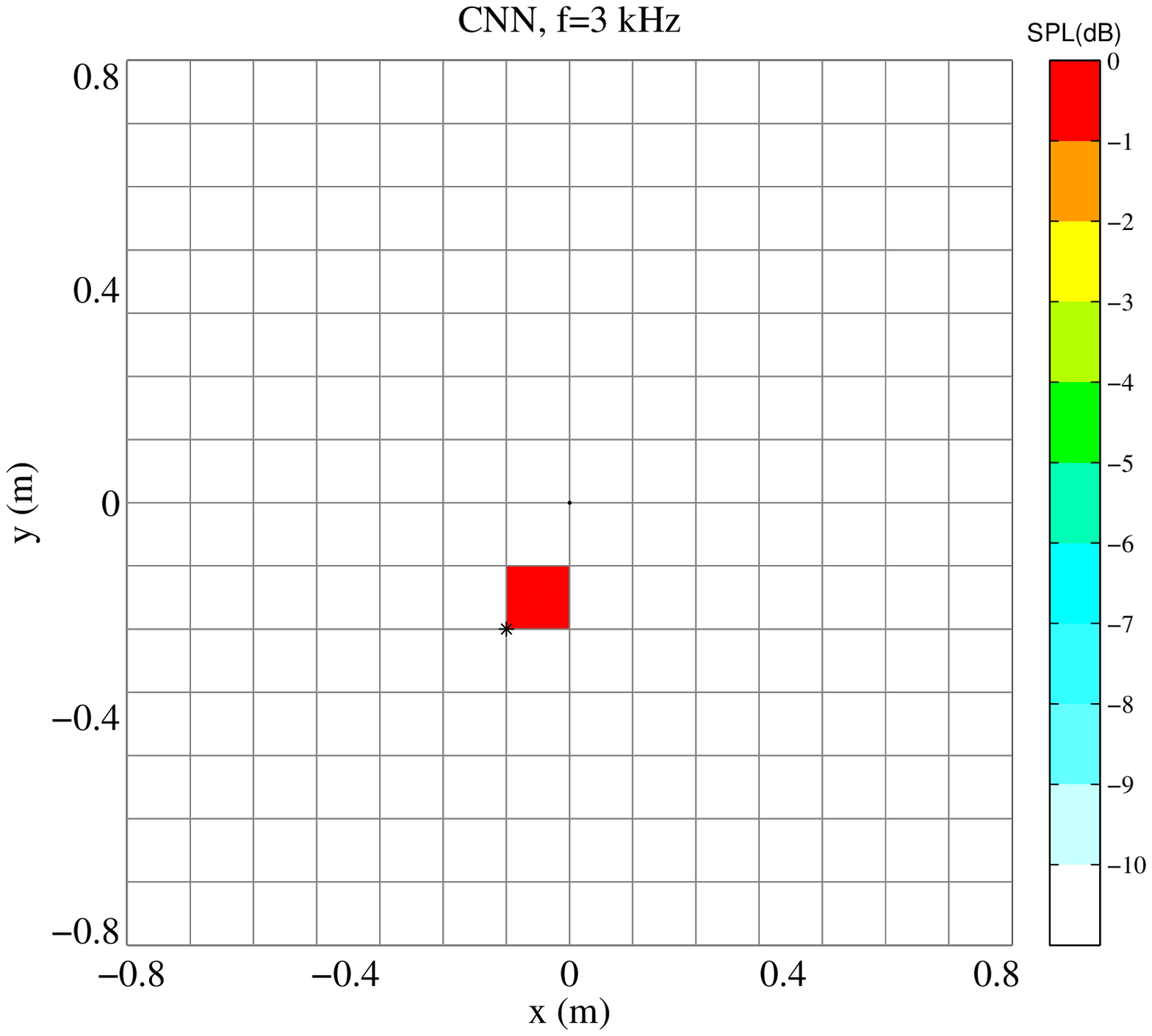}}\\
\caption[]{$f$=3 kHz. Black cross symbols, positions of synthetic point sources. The first line, three dispersed sources far apart; the second line, two of three sources are adjacent. The first column, conventional beamforming; the second column DAMAS; the third column, CNN.}
\label{fig:f3}
\end{figure}

Unfortunately CNN test accuracy decreases as the frequency decreases.
CNN test accuracies are only 83\% and 60\% at $f$=5 kHz and 3 kHz, respectively.
After checking these incorrect examples at $f$=3 kHz, at least one sound source point is located at the edge of grid in lots cases.
These incorrect examples are the same as those in $f$=8 kHz.
However there are also lots incorrect examples where sound sources are adjacent.
Fig. \ref{fig:f3} shows the reconstruction results at $f$=3 kHz for the two given sound distributions in Fig. \ref{fig:f8}.
In the first given sound distribution with three dispersed sources far apart, conventional beamforming cannot separate these three sources, because resolution of conventional beamforming is inversely proportional to frequency according their relationship \cite{Hald-2004a}, $R= \dfrac{1.22}{\text{cos}(\alpha/2)^3}\dfrac{zc}{Df}$, where $R$ is resolution of conventional beamforming and $c$ is sound velocity. $R$ and grid spacing $\Delta x$ are also listed in Table \ref{tab:CNN_accuracy}.
For this sound distribution, both CNN and DAMAS can reconstruct accurately the source distribution.
In the second given sound distribution with two adjacent sources, conventional beamforming cannot separate sources unexpectedly, while DAMAS can reconstruct accurately the source distribution.
However CNN losses its accuracy, at this given sound distribution.

With regard to computing speed in applications, CNN is as fast as conventional beamforming, and is significantly faster than DAMAS.

\begin{table}[htbp]
\centering
\caption{Parameters with frequency.}
\label{tab:CNN_accuracy}
\begin{tabular}{cccc}\hline
 $f$ & 8 kHz & 5 kHz & 3 kHz \\\hline
CNN test accuracy & 98$\%$ & 83$\%$ & 60$\%$ \\
$R$ & 0.3757&0.6012 &1.0019  \\
$\Delta x$ & 0.1183 & 0.1183 & 0.1183\\\hline
\end{tabular}
\end{table}

%

\clearpage
\section{Discussion and Conclusion}\label{sec:discussion}
In this paper, CNN a kind of deep learning as an alternative algorithm is preliminarily applied to phased microphone arrays for sound source localization.
To the best knowledge of the authors, this paper is first work so far that applies deep learning to phased microphone array for sound source localization.

Through preliminary investigation, at high frequency CNN can reconstruct the sound localizations with excellent spatial resolution as good as DAMAS, within a very short time as short as conventional beamforming.
This exciting result means that CNN almost perfectly finds source distribution from cross-spectral matrix without given propagation function in advance.
This preliminary investigation makes CNN has encouraging prospects for applications with unknown propagation function, and thus CNN deserves to be further explored as a new algorithm.

One point the authors want to emphasize here is that CNN is definitely not data fitting.
For three uniform sound sources randomly distributed in 225 grids, there are $C_{225}^3$=1.873$\times$10$^6$ possibilities.
In the applications, the number of training data is 3.2$\times$10$^4$, only 1.71\% of all the possibilities.

About the CNN investigation and optimization, the questions are still open and needed to investigate in the future, such as: (i) What's the dynamic range of CNN? (ii) How many layers are most suitable for a give data set? (iii) How many kernel number and size are needed? (iv) How big are training data? (v) What is the uncertainty of CNN predictions? (vi) How to improve the accuracy at low frequency? (vii) How to improve the accuracy when sound sources located at edge of grid?



The main challenge of CNN is that large amount of training data are required.
Especially in the applications with unknown propagation function, reliable training data can only be accumulated through a large number of experiments with a process that takes a lot of time and money.



\bibliographystyle{model1-num-names}


\end{document}